\documentclass[epj]{svjour}
\pdfoutput=1
\usepackage{graphicx}
\input{epsf}
\usepackage{amssymb}

\begin{document}

\title{Delocalization of two interacting particles 
in the two-dimensional Harper model}

\author{
K.M. Frahm
%\inst{1}
\and
D.L. Shepelyansky
%\inst{1}
}
\institute{
Laboratoire de Physique Th\'eorique du CNRS, IRSAMC, 
Universit\'e de Toulouse, UPS, 31062 Toulouse, France
}

\titlerunning{Delocalization of two interacting particles in the 
2D Harper model}
\authorrunning{K.M.Frahm  and D.L.Shepelyansky}

\abstract{We study the problem of two interacting particles in a 
two-dimensional quasiperiodic potential of the Harper model. 
We consider an amplitude of the quasiperiodic potential such that 
in absence of interactions all eigenstates are 
exponentially localized while the two interacting particles
are delocalized showing anomalous subdiffusive spreading over the lattice
with the spreading exponent $b \approx 0.5$ instead of a usual diffusion with 
$b=1$. This spreading is stronger than in the case of a correlated disorder 
potential with a one particle localization length as 
for the quasiperiodic potential. 
At the same time we do not find signatures of ballistic FIKS pairs
existing for two interacting particles in the one-dimensional Harper model.
}


%\date{\today}
\date{Dated: October 5, 2015}

\maketitle

\section{Introduction}
\label{sec1}
The Harper problem describes the quantum dynamics of an electron
in a two-dimensional potential (2D) in a perpendicular 
magnetic field \cite{harper}. It can be reduced to the Schr\"odinger
equation on a discrete quasiperiodic one-di\-mensional (1D) lattice.
This system has fractal spectral properties \cite{hofstadter}
and demonstrates a Metal-Insulator Transition (MIT), 
established by Aubry and Andr\'e \cite{aubry}. The MIT takes place when
the amplitude $\lambda$ of the quasiperiodic potential (with hopping 
being unity) is changed from $\lambda <2$ (metallic phase) 
to $\lambda>2$ (insulator phase).
A review of the properties of the Aubry-Andr\'e model can be found 
in \cite{sokoloff} and the mathematical prove of the MIT is given in 
\cite{lana1}.

The investigation of interaction effects between particles in the 
1D Harper model was started in \cite{dlsharper} with the case of 
Two Interacting Particles (TIP). 
It was found that the Hubbard interaction can create TIP localized states in
the noninteracting metallic phase. Further studies also demonstrated the localization effects
in presence of interactions \cite{barelli,orso}. 
This trend was opposite to the TIP effect in disordered systems
where the interactions increase the TIP localization length in 1D or even
lead to delocalization of TIP pairs for dimensions $d \geq 2$ 
\cite{dlstip,imry,pichard,frahm1995,vonoppen,borgonovi,dlsmoriond,frahmtip,dlscoulomb,lagesring}.

Thus the results obtained in \cite{flach}
on the appearance of delocalized TIP pairs in the 1D Harper model,
for certain particular values of interaction strength and energy, 
in the regime, when all one-particle states are exponentially localized,
is really striking. In \cite{flach} the delocalization of TIP
appears at a relatively strong interaction 
being    the reason why this effect was missed in previous studies.
The recent advanced analysis \cite{fiks1d}
showed that so called Freed by Interaction Kinetic States (FIKS)
appear at various irrational magnetic flux values
being ballistic or quasi-ballistic over the whole system size $N$
used in numerical simulations (up to $N=10946$).
At certain flux values the FIKS pairs appear even at
a moderate Hubbard interaction $U=1.75$ (hopping is taken as $t=1$),
also the effect of FIKS pairs becomes stronger 
for long range interactions \cite{fiks1d}. Up to 12\%
from an initial state, with TIP being close to each other,
can be projected on the FIKS pairs escaping ballistically to infinity 
\cite{fiks1d}.
This observation points to possible significant applications of FIKS pairs
in various physical systems and shows the importance of further 
investigations of the FIKS effect.
Indeed, as shown in \cite{fiks1d}, the recent experiments with cold atoms 
on quasiperiodic lattices \cite{roati,modugno,bloch} should be able to   
detect FIKS pairs in 1D.

For the TIP effect in disordered systems the dimension plays an important role
\cite{imry,borgonovi,dlsmoriond,dlscoulomb,lagesring} and it is clear that 
it is important to study the FIKS effect in higher dimensions. We start 
these investigations here for the two-dimensional (2D) Harper model 
where the (noninteracting) eigenstates are given by the product of two 
1D Harper (noninteracting) eigenstates so that the MIT 
position for noninteracting states is clearly defined at $\lambda =2$. 
We note that 2D quasiperiodic lattices of cold atoms have been realized 
in recent experiments (even if the second dimension was a repetition 
of 1D lattices) \cite{bloch2d} so that there are new possibilities to 
investigate the FIKS effect with cold atoms when the interaction is taken 
into account. 

The paper is composed as follows: the model description is given in Section 2,
the main  results are presented in Section 3, discussion of results
is given in Section 4. High resolution figures and additional data
are available at the web site \cite{webpage}.

\section{Model description}
\label{sec2}

We consider particles in a 2D lattice 
of size $N_1\times N_2$, $0\le x<\! N_1$ and $0\le y <\! N_2$. The 
one-particle Hamiltonian $h^{(j)}$ for particle $j$ is given by:
\begin{eqnarray}
\label{eq_h1}
h^{(j)}&=&T^{(j)}+V^{(j)},\\
\label{eq_T1}
T^{(j)}&=&-\sum_{x,y} \Bigl(|x,y\!>_j\,<\!x+1,y|_j\\
\nonumber
&&\qquad\quad +|x,y\!>_j\,<\!x,y+1|_j\Bigr)+h.~c.,\\
V^{(j)}&=&\sum_{x,y} \Bigl[V_1(x-x_0)+V_2(y-y_0)\Bigr]|x,y\!>_j\,<\!x,y|_j.
\label{eq_V1}
\end{eqnarray}
The point $(x_0,y_0)=(N_1/2,N_2/2)$ is the 
``center point'' of the lattice and the offsets $x-x_0$ or $y-y_0$ 
in the arguments of $V_1$ ensure that the potential has locally the 
same structure for the region close to the center point when varying 
the system size $N_1\times N_2$. 
The kinetic energy $T^{(j)}$ is given by the standard tight-binding 
model in two dimensions with hopping elements $t=-1$ linking nearest neighbor 
sites with periodic boundary conditions, i.~e. $x+1$ (or $y+1$) in 
(\ref{eq_T1}) is taken modulo $N_1$ (or $N_2$). 
Note that the potential is of the form 
\begin{equation}
\label{eq_pot_form}
V(x,y)=V_1(x-x_0)+V_1(y-y_0)
\end{equation}
%$V(x,y)=V_1(x-x_0)+V_2(y-y_0)$ 
where 
$V_1(x)$, $V_2(y)$ are effective one-dimensional potentials. 
In this work we study essentially the quasiperiodic case 
with $V_1(x)=\lambda_x\cos(\alpha x+\beta)$,
$V_2(y)=\lambda_y\cos(\alpha y+\beta)$ and here mostly 
$\lambda_x=\lambda_y=\lambda=2.5$. Furthermore we choose  
$\alpha=2\pi(\sqrt{5}-1)/2\approx 0.61803$ as the golden ratio 
and $\beta=1/\sqrt{2}$. For these parameters the 
one-dimensional eigenfunctions (with the $V_1$ potential) are localized 
with a one-dimensional localization length 
$\ell=1/\log(\lambda/2)\approx 4.48$ (see e.g. \cite{sokoloff,fiks1d}). 
For the purpose of comparison we also study the disorder case with a 
random potential $V_1(x)$ uniformly distributed in $[-W/2,W/2]$ and the 
same random realization 
for $V_2(y)$. For this case we choose $W=5$ corresponding to the localization 
length $\ell\approx 105/W^2\approx 4.2$ which is quite close to the 
localization length of the quasiperiodic case for $\lambda=2.5$. 
The particular structure of $V$ implies that for both cases 
the eigenfunctions of $h^{(j)}$ 
are products of one-dimensional localized eigenstates 
in $x$ and $y$ with the potential $V_1(x-x_0)$ or $V_2(y-y_0)$. 

We note that for the disorder case the potential $V(x,y)$ is due to the 
particular sum structure in Eq. (\ref{eq_V1}) 
very different from the standard Anderson 
two-dimensional disorder model. In the latter case $V(x,y)$ would be 
independent random variables for each value of $(x,y)$ 
while in our case $V(x,y)$ is a sum of two one-dimensional disorder potentials 
providing certain spacial correlations in the potential 
which are crucial for the value of the quite small localization length. 

We now consider two interacting particles, each of them submitted to the 
one-particle Hamiltonian $h^{(j)}$, and coupled by an interaction potential 
$U(x_1,y_1,x_2,y_2)$ which has a non-vanishing value $U$ only for 
$|x_1-x_2|<\!U_R$ and $|y_1-y_2|<\!U_R$ \cite{U_boundary}.
Here $U$ denotes the interaction strength 
and $U_R$ is the interaction range. The total two particle Hamiltonian is 
given by 
\begin{equation}
\label{ham_tot}
H=h^{(1)}+h^{(2)}+\hat U
\end{equation}
where $\hat U$ is the interaction operator in the two-particle 
Hilbert space with diagonal entries $U(x_1,y_1,x_2,y_2)$. 
In this work we consider two cases with 
$U_R=1$, corresponding to Hubbard on-site-interaction, and $U_R=2$ 
corresponding to a short range interaction with 9 neighboring sites coupled 
by the interaction. 

The eigenfunctions of $H$ are either symmetric with respect to particle 
permutation (boson case) or anti-symmet\-ric (fermion case) corresponding 
to a decomposition of the Hilbert space in a boson- and fermion-subspace. 
However, in this work we prefer to work on the complete space (of dimension 
$N_1^2\,N_2^2$) due to the employed numerical method to determine the 
time evolution of the wave function. The evolution is described by
the  time-dependent Schr\"odinger equation (with $\hbar=1$)
\begin{equation}
\label{eq_schroedinger}
i\frac{\partial}{\partial t}\,|\psi(t)\!>=H\,|\psi(t)\!>.
\end{equation}
The symmetry of the state $|\psi(t)\!>$ is simply fixed by the 
symmetry of the initial condition which is conserved by 
the Schr\"odinger equation and which we choose
\begin{equation}
\label{eq_initial}
|\psi(0)\!>=|x_0,y_0\!>_1\,|x_0,y_0\!>_2
\end{equation}
corresponding to both particles being localized on the same center point 
with $x_0=N_1/2$ and $y_0=N_2/2$. 

As already noted, in absence of the interaction, i.~e. $U=0$, the 
eigenstates are localized with a typical localization length $\ell$ (in each 
direction). Thus, our aim is to study if interaction leads to 
a delocalization of TIP 
during the time evolution or to some kind 
of diffusion of TIP in coordinate or Hilbert space. 

To solve (\ref{eq_schroedinger}) numerically we write $H=H_x+H_p$ as a sum of 
two parts which are either diagonal in position space 
$H_x=V^{(1)}+V^{(2)}+\hat U$ or in momentum space $H_p=T^{(1)}+T^{(2)}$
and evaluate the solution of (\ref{eq_schroedinger}) as:
\begin{equation}
\label{eq_solution}
|\psi(t)\!>=\exp(-iHt)\,|\psi(0)\!>
\end{equation}
using the Trotter formula approximation: 
\begin{eqnarray}
\label{eq_trotter}
\exp(-iHt)&\approx& (O_p\,O_x)^{t/\Delta t},\\
\nonumber 
O_p&=& \exp(-iH_p\,\Delta t),\quad 
O_x= \exp(-iH_x\,\Delta t)
\end{eqnarray}
with two unitary operators $O_p$ and $O_x$. 
The integration time step $\Delta t$ 
is supposed to be small as compared to typical inverse 
energy scales and the value of $t$ is chosen such that 
$t/\Delta t$ is integer. Formally, Eq. (\ref{eq_trotter}) becomes exact 
in the limit $\Delta t\to 0$. However, a finite value of $\Delta t$ implies
a modification of the Hamiltonian with $H\to \tilde H$ with $\tilde H$ 
defined by $O_p\,O_x=\exp(-i\tilde H\,\Delta t)$ and related to $H$ 
by a power law expansion in $\Delta t$ where the corrections are 
given as (higher order) commutators between $H_x$ and $H_p$. 
In this work we choose the value $\Delta t=0.1$ but we have verified for 
certain parameter values that the results presented below do not change 
significantly if compared with $\Delta t=0.05$. The efficiency and stability 
of this type of integration methods have been demonstrated 
in \cite{dlstip,fiks1d,danse,garcia}.

The operators $O_x$ and $O_p$ are either diagonal in position representation 
or momentum representation. In order to evaluate (\ref{eq_solution}) 
using (\ref{eq_trotter}) we first apply the operator $O_x$ to 
the initial state given in position representation which can be done 
efficiently 
with $N_{\rm tot}=N_1^2\,N_2^2$ operations by multiplying the eigenphases of 
$O_x$ to each component of the state. Then the state is transformed to 
momentum representation using a fast Fourier transform in the 
four dimensional configuration space (corresponding to two particles in 
two dimensions) with help of the library FFTW \cite{fftw} which requires about 
$N_{\rm tot}(\log N_1+\log N_2)$ operations. At this point we can 
efficiently apply the operator $O_p$ to the states, again by multiplying 
the eigenphases to each component of the state and finally we apply 
the inverse Fourier transform to come back in position representation. 
The eigenphases of $O_x$ and $O_p$ can be calculated and stored in advance. 

We determine the time evolution of $|\psi(t)\!>$ using 
Eq. (\ref{eq_trotter}) for different square and rectangular geometries 
with system sizes up to $128\times 128$ (i.~e. $N_1=N_2=128$) or 
$1024\times 8$ (i.~e. $N_1=1024$, $N_2=8$). 
At $N_1=N_2=128$ the Hilbert space of the whole system becomes as large as
$N_H = N_1^4 \approx 2.7 \times 10^8$.
In order to analyze the 
structure of the TIP state we introduce different quantities and 
densities described below.

First let us denote by
\begin{equation}
\label{eq_wavefun}
\psi(x_1,y_1,x_2,y_2)=<\!x_1,y_1|_1<\!x_2,y_2|_2\,\psi\!>
\end{equation}
the (non-symmetrized) two particle wave function and for simplicity we 
omit the argument for the time dependence. Then the one-particle density 
$\rho_1(x,y)$ in 2D is defined as
\begin{equation}
\label{eq_onepartdens}
\rho_1(x,y)=\sum_{x_2,y_2}|\psi(x,y,x_2,y_2)|^2.
\end{equation}
We note that the normalization of the state $|\psi\!>$ 
implies $\sum_{x,y}\rho_1(x,y)=1$. 
Using this one-particle density we define the variance with 
respect to the center point $(x_0,y_0)$ by
\begin{equation}
\label{eq_variance}
\langle r^2\rangle=\sum_{x,y} \Bigl[(x-x_0)^2+(y-y_0)^2]\,\rho_1(x,y)
\end{equation}
and also the inverse participation ratio (IPR) ``without center'' by:
\begin{equation}
\label{eq_inversepart}
\xi_{\rm IPR}=\frac{\left[\sum_{(x,y)\in S} \rho_1(x,y)\right]^2}
{\sum_{(x,y)\in S} \rho_1^2(x,y)}
\end{equation}
where the sums run over the set 
\begin{equation}
\label{eq_set}
S=\Bigl\{(x,y)\ \Big|\ |x-x_0|> N_1/10\ ,\ |y-y_0|> N_2/10\Bigr\}
\end{equation}
containing only lattice sites $(x,y)$ outside the center rectangle of 
(linear) size 
20\% around the center point $(x_0,y_0)$. This kind of definition for 
the IPR allows to detect a particular kind of 
partial delocalization where only a small fraction of probability diffuses to 
large distances with respect to the center point while the remaining 
probability stays strongly localized close to the center point. 
This quantity was already used with success in our studies of FIKS pairs in 
\cite{fiks1d} for the 1D TIP Harper problem. 
Using the standard definition for the 
IPR (where $S$ would be the set of {\em all} 
lattice sites) allows only to detect a strong delocalization of the 
full probability. For the variance $\langle r^2\rangle$ the contribution
of the probability at the initial state is not so pronounced
and thus we compute this quantity for the whole lattice.

We furthermore introduce the following densities 
\begin{eqnarray}
\label{eq_rhox}
\rho_x(x)&=&\sum_y\rho_1(x,y),\\
\label{eq_rhoy}
\rho_y(y)&=&\sum_x\rho_1(x,y),\\
\label{eq_rhoxx}
\rho_{xx}(x_1,x_2)&=&\sum_{y_1,y_2}|\psi(x_1,y_1,x_2,y_2)|^2,\\
\label{eq_rholin}
\rho_{\rm lin}(s)&=&\sum_{\scriptsize
\begin{array}{c}
x,y  \\
s=|x-x_0|+|y-y_0| \\
\end{array}} \rho_1(x,y). 
\end{eqnarray}
The density $\rho_x(x)$ (or $\rho_y(y)$) 
is simply the one-particle density integrated 
over the $y$-direction (or $x$-direction). 
$\rho_{xx}(x_1,x_2)$ is the two particle density 
integrated over both $y$-directions giving information about the spatial 
correlations of both particles in $x$-direction. Here
$\rho_{\rm lin}(s)$ is the linear density obtained from the one-particle 
density by summing over all sites with same (1-norm)-distance 
$s=|x-x_0|+|y-y_0|$ from the center point and is well defined 
for $0\le s< (N_1+N_2)/2$. This density is similar in 
spirit to a radial density obtained by integrating over all points with the same 
distance from the center point. However, using the 1-norm (and not 
the Euclidean 2-norm) to measure the distance is both more convenient for 
the practical calculation and actually physically more relevant for the case
where $\rho_1(x,y)\sim \exp[-(|x-x_0|+|y-y_0|)/l]=\exp(-s/l)$ is 
similar to a product of two exponentially localized 
functions in $x$ and $y$ with the same localization length $l$. 

\section{Time evolution results}
\label{sec3}

\begin{figure}
\begin{center}
\includegraphics[width=0.48\textwidth]{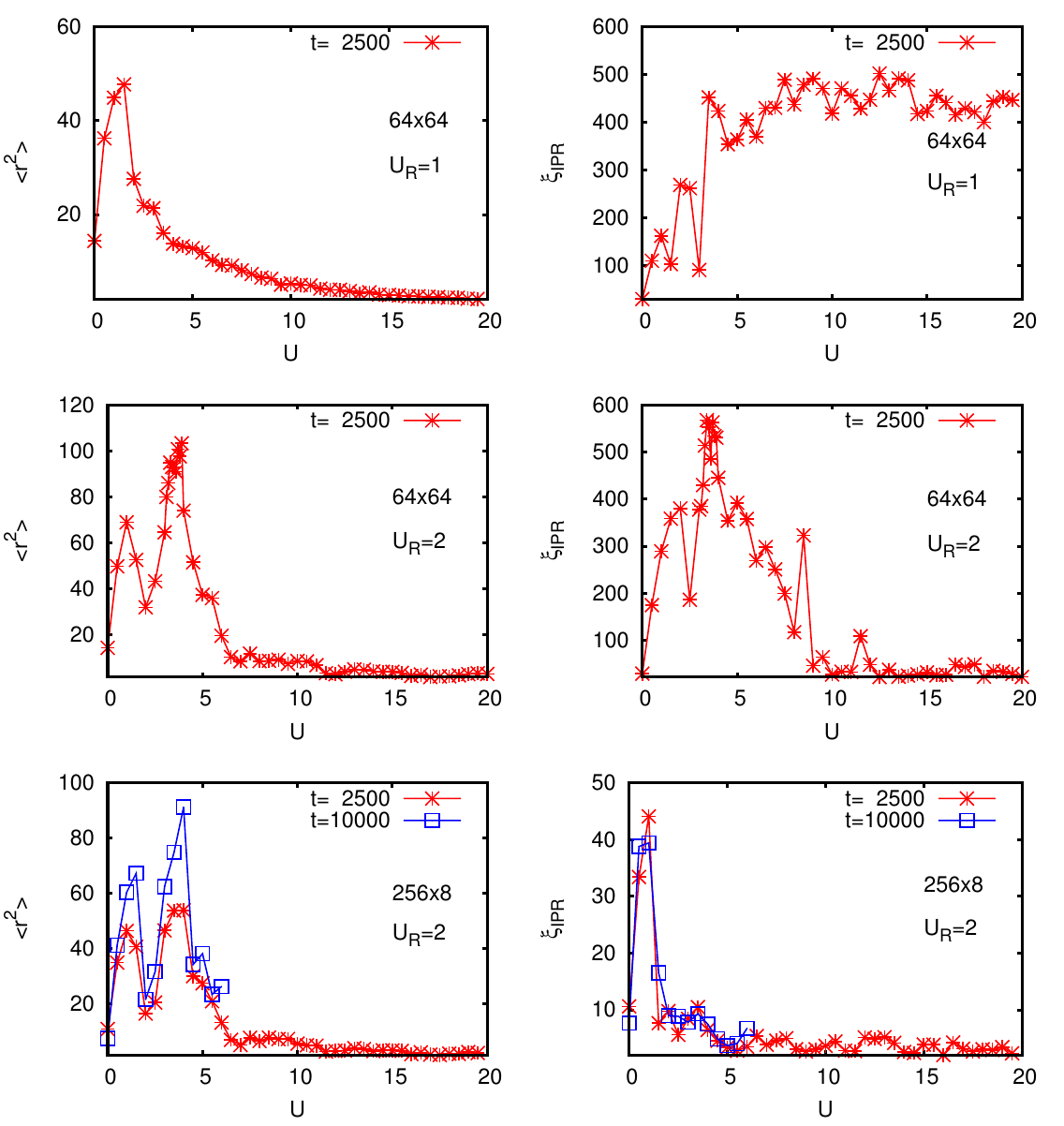}
\caption{Variance $\langle r^2\rangle$ (left column) 
and IPR without center $\xi_{\rm IPR}$ (right column) 
versus interaction strength $U$ for $0\le U<20$
for 2D quasiperiodic potential ($\lambda=2.5$). 
The top panels correspond to Hubbard interaction 
($U_R=1$) and the square geometry $N_1=N_2=64$; the center panels 
correspond to $U_R=2$ for $N_1=N_2=64$; the  
bottom panels correspond to $U_R=2$ with rectangular geometry 
$N_1=256,\ N_2=8$. In all panels the iteration time is $t=2500$ except 
for the two bottom panels where additional data points for $t=10000$ and 
$0\le U\le 6.0$ are shown. 
}
\label{fig1}
\end{center}
\end{figure}

\begin{figure}
\begin{center}
\includegraphics[width=0.48\textwidth]{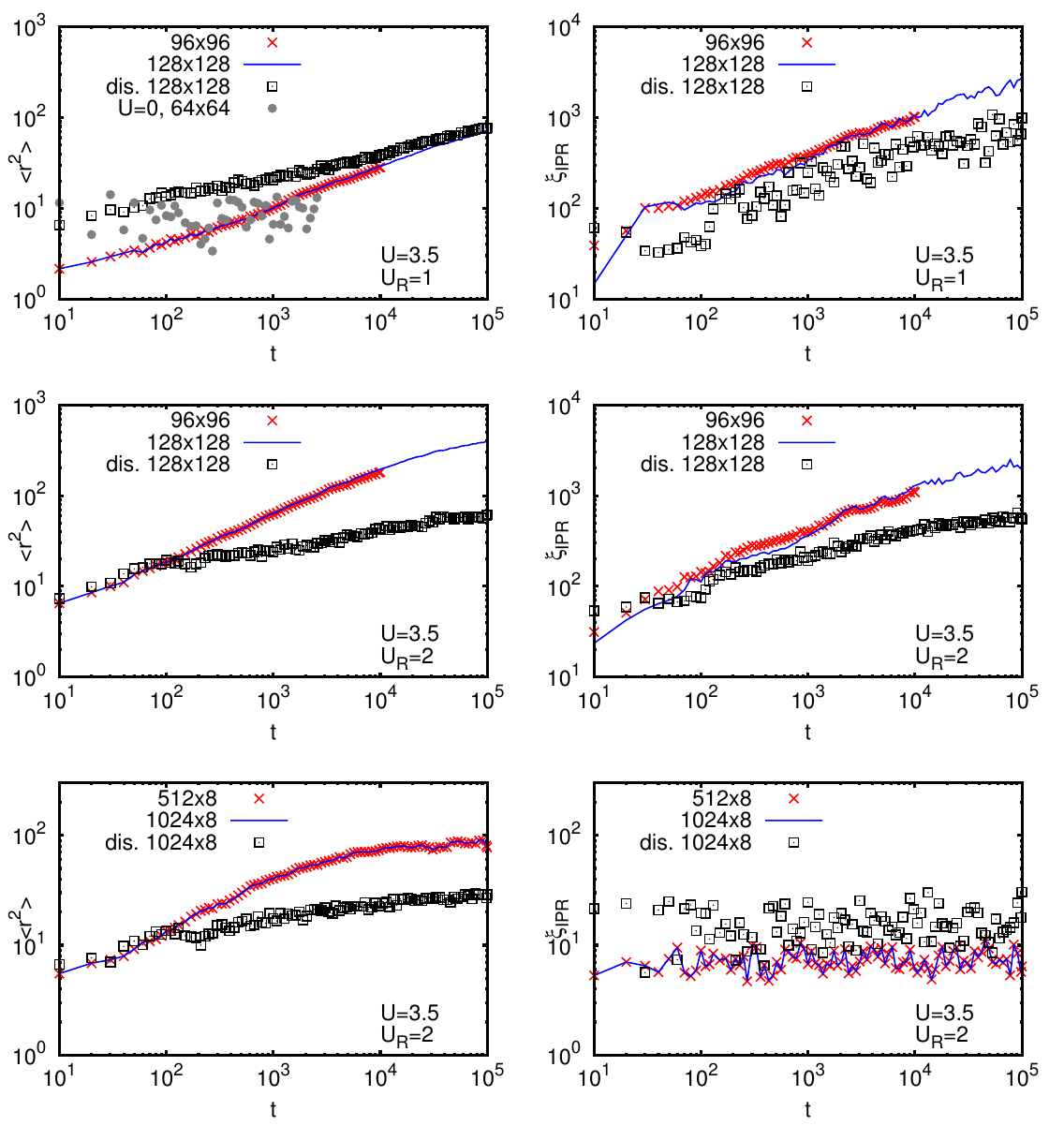}
\caption{Variance $\langle r^2\rangle$ (left column) 
and IPR without center $\xi_{\rm IPR}$ (right column) 
versus iteration time $10\le t\le 10^5$ in a double logarithmic scale. 
The two top panels correspond to Hubbard interaction with interaction range 
$U_R=1$, the square geometry $N_1=N_2=128$ (blue curve) or $N_1=N_2=96$ 
(red crosses for $10\le t\le 10^4$). In the top left panel also 
data for $U=0$, $N_1=N_2=64$, $10\le t\le 2500$ (grey points) are shown. 
The two center panels 
correspond to $U_R=2$ and the same geometries as in 
the top panels. 
The two bottom panels correspond to $U_R=2$ 
with rectangular geometries $N_1=1024$, $N_2=8$ (blue curve) 
and $N_1=512$, $N_2=8$ (red crosses). 
Furthermore in all panels data for a random disorder potential ($W=5$) 
of the particular form (\ref{eq_pot_form}) (see text) 
and same panel values $N_1, N_2$ are shown by black squares. 
In all panels the interaction strength is $U=3.5$. 
All curves and symbols, except the black squares, correspond to the 2D Harper
model at $\lambda=2.5$. The number of shown 
data points is artificially reduced to increase the visibility. 
}
\label{fig2}
\end{center}
\end{figure}

As in \cite{fiks1d} we first determine the most promising values of 
the interaction strength $U$ by computing $\langle r^2\rangle$ 
and $\xi_{\rm IPR}$ at a certain large $t$. Here we use a moderate 
system size since computations should 
be done for many values of $U$ at $U_R=1$ (Hubbard interaction) and
$U_R=2$ (9 nearest sites coupled on a square lattice). 
The results are presented in Fig.~\ref{fig1}. 
We see that there are regions of $U$  where  
the values of $\langle r^2\rangle$
are by a factor $4-10$  larger than in the case of $U=0$ where
$\langle r^2\rangle \approx 10$ (see Fig.~\ref{fig2}). 
However, in contrast to the 1D TIP Harper model \cite{fiks1d}
there are no sharp peaks in $U$ except maybe at $U=3.5$ for $U_R=2$. 
In the following, we choose this value for
a more detailed analysis at larger sizes $N_1, N_2$ and larger times $t$. 
However, we have also studied some other $U$ values, e.~g. $U=6$ 
with qualitatively similar results but typically with less delocalization than 
the most interesting value $U=3.5$. 

In Fig.~\ref{fig2}, we show for $U=3.5$, the two values $U_R=1$ and 
$U_R=2$ and different geometries the time dependence 
of $\langle r^2\rangle$ and $\xi_{\rm IPR}$. 
All the cases with a square geometry $N_1=N_2$ show an 
unlimited growth of these two 
quantities up to largest times $t=10^5$ reached in our numerical simulations.
For the Hubbard case at $U_R=1$ the system size is sufficiently large 
to avoid saturation effects due the finite system size 
and the change of size from $N_1=N_2=96$
to $128$ does not affect the values of $\langle r^2\rangle$ and  $\xi_{\rm IPR}$ at $U=3.5$.
For $U_R=2$ we have larger values of $\langle r^2\rangle$ and  $\xi_{\rm IPR}$
and it is clear that the size $N_1=N_2=96$ is sufficiently large only up 
to $t \approx 10^4$ while for $N_1=N_2=128$ the size is sufficient 
only up to $t \approx 3 \times 10^4$ with a finite size induced 
saturation of growth for $3  \times 10^4 < t \leq 10^5$.

In a drastic contrast with the 
1D case \cite{fiks1d} we observe only a subdiffusive
growth of $\langle r^2\rangle \propto t^{b_1}$ and 
$\xi_{\rm IPR} \propto t^{b_2}$ with time. 
The power law fits of the data used in Fig.~\ref{fig2} provide the values: 
$b_1=0.438 \pm 0.004$, $b_2=0.503 \pm 0.007$ for $U_R=1$;  
$b_1=0.521 \pm 0.002$, $b_2=0.506 \pm 0.009$ for $U_R=2$
for the range $100 \leq t \leq 10^5$ at $N_1=N_2=128$.

For comparison, we also present in Fig.~\ref{fig2} 
the same quantities for the case of the particular 
disordered potential described in Section \ref{sec2}. For this we use 
the same interaction strength $U=3.5$ and 
the disorder parameter $W=5$ which gives approximately the same 
localization length in 1D as for the 1D Harper model at $\lambda=2.5$
(however, for the usual 2D Anderson model we would have a significantly larger
value of the one-particle IPR  $\xi \approx 150$, 
see e.g. Fig.~2 in \cite{lagesbls}).
For $U_R=2$ and $t>10^2$ both the absolute values and the growth 
rates of $\langle r^2\rangle$ and $\xi_{\rm IPR}$ for the disorder case are 
significantly lower as compared to the 2D Harper model. 
For $U_R=1$ the disorder values of the variance are above 
the variance values of the 2D Harper model, for the time interval 
$10\leq t\leq 10^5$ shown in the figure, but the curve for the Harper case 
has a stronger growth rate (larger slope). 

Actually, according to Fig.~\ref{fig2}  the two  curves for $\langle r^2\rangle$ seem to 
intersect at a certain time $t_{\rm int}$ and therefore we expect the 
variance of the 2D Harper model to become stronger than the variance of the 
disorder case for $t>t_{\rm int}$. From  the figure it seems  
that $t_{\rm int}$ is close or slightly below $10^5$ but this 
is only due to the rather thick data points and the logarithmic scale. 
A careful analysis of the data (higher resolution figure and more precise 
extrapolation of both curves using power law fits for $10^4\le t\le 10^5$) 
shows that the intersection point is likely to be close to the value 
$t_{\rm int}\approx 2.4\times 10^5$. 
For $U_R=1$, the another quantity $\xi_{\rm IPR}$ for the disorder case is
clearly below the curve of the Harper 
model. Our interpretation is that apparently for TIP in the disorder case there is 
a relative strong initial spreading at short times and a modest 
length scale but for a strong weight of the wave packet while for the 
Harper case there is a slower but long range delocalization 
for a smaller weight of the wavepacket which is better visible from 
the IPR $\xi_{\rm IPR}$ without the center rectangle. 
(This kind of ``long range small weight'' delocalization was also found 
for the FIKS pairs of the TIP 1D Harper model \cite{fiks1d} but there the 
growth rate is actually ballistic, corresponding to power law exponents 
$b_{1,2}\approx 2$, and not sub-diffusive.)

The lower growth rate for the disorder case at both 
values of $U_R$ is also clearly confirmed by 
the power law fits which provide (for the same time and size ranges as 
for the Harper case) the exponents: 
$b_1= 0.218 \pm 0.005$, $b_2= 0.404 \pm  0.035$ for $U_R=1$
and $b_1= 0.181 \pm 0.007$, $b_2= 0.302 \pm  0.009$ for $U_R=2$.

In Fig.~\ref{fig2} we also consider the case of two rectangular geometries 
with $N_1=1024$ or $N_1=512$ and $N_2=8$. In this case there is a clear
saturation of growth of the considered variables independent of the system 
size. These data show that for $N_2 \sim \ell$ we have a localization of 
TIP in the quasi-1D Harper model at the considered 
interaction strength. However, this result does not exclude
the possibility of appearance of FIKS pairs in the quasi-1D limit at
other interaction values, even if our preliminary tests
indicate similar localization results.

\begin{figure}
\begin{center}
\includegraphics[width=0.48\textwidth]{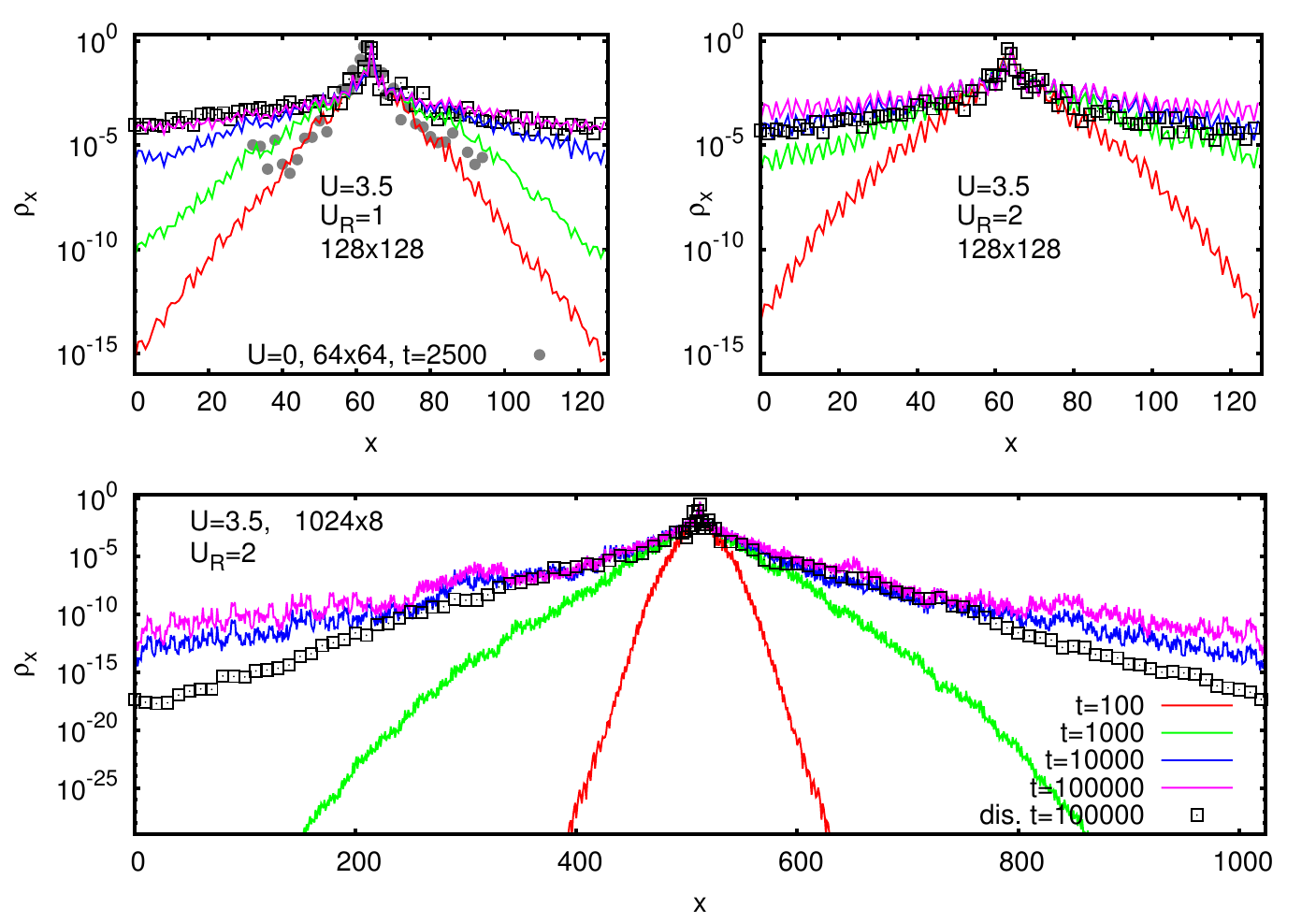}
\caption{Density $\rho_x(x)$ versus $x$ in a semilogarithmic representation 
for different values of interaction 
range, geometry and iteration times at $U=3.5$. The color labels 
shown in the bottom right corner of the bottom panel apply to all three 
panels: $t=100$ (red curve), $t=1000$ (green curve), $t=10000$ (blue curve), 
$t=100000$ (pink curve), disorder potential ($W=5$) 
for $t=100000$ (black squares). 
In the top left panel also 
data for $U=0$, $N_1=N_2=64$, $t\le 2500$ (grey points) are shown (with 
center point shifted from 32 to 64). 
All curves, except the black squares, correspond to the quasiperiodic 
potential ($\lambda=2.5$).
For grey points and black squares the number of shown 
data points is artificially reduced to increase the visibility.}
\label{fig3}
\end{center}
\end{figure}

The time evolution of the projected one-particle probability distribution
$\rho_x(x)$ is shown in Fig.~\ref{fig3}. For the square geometry
$N_1=N_2=128$ the width of the distribution is growing with time
and it becomes practically flat at maximal times $t=10^5$ for 
both values $U_R=1$ or $U_R=2$. 
In the case of disorder we have also a significant
spreading of probability over lattice sites which is somewhat
comparable with those of the 2D Harper case.
For the rectangular geometry we have a significantly larger probability 
on the tails
for the 2D Harper model as compared to the disorder case. This is in agreement
with the data for  $\langle r^2\rangle$ in Fig.~\ref{fig2} (bottom left panel).

\begin{figure}
\begin{center}
\includegraphics[width=0.48\textwidth]{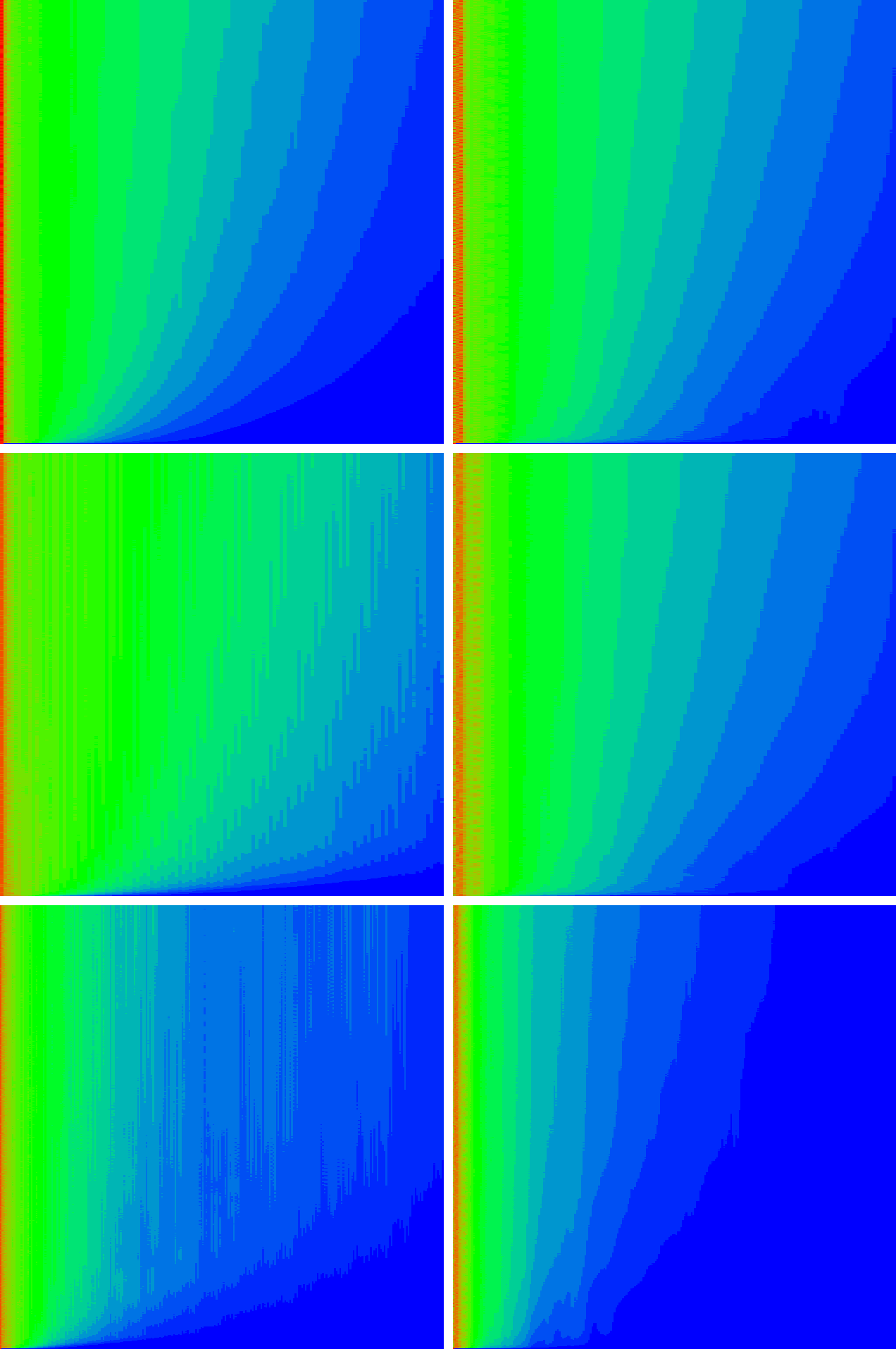}
\caption{Density plot of the time evolution of the linear density 
$\rho_{\rm lin}(s)$. 
The vertical axis corresponds to the iteration time $0\le t\le 10000$ 
and the horizontal axis corresponds to $0\le s < (N_1+N_2)/2$. 
The left column corresponds to the quasiperiodic potential ($\lambda=2.5$) 
and the right column to the disorder case ($W=5$). 
All panels correspond to the interaction strength $U=3.5$. 
Top (center) panels correspond 
to $U_R=1$ ($U_R=2$) and the square geometry $N_1=N_2=128$. 
Bottom panels correspond 
to $U_R=2$ and the rectangular geometry $N_1=512$, $N_2=8$. 
The color codes of the density plot correspond to red for maximum, 
green for medium and blue for minimum values.}
\label{fig4}
\end{center}
\end{figure}

These results show that there are no ballistic type FIKS pairs 
propagating through the whole system as it was the case
for TIP in the 1D Harper model \cite{flach,fiks1d}. Such a conclusion
is confirmed by the analysis of the time evolution of the 
linear density $\rho_{\rm lin}(s)$ defined in (\ref{eq_rholin})
as shown in Fig.~\ref{fig4}. The typical width of this density 
does not increase linearly in time in contrast to the 1D Harper case 
(see e.g. Fig.~3 in \cite{fiks1d}) and 
we have in Fig.~\ref{fig4} (for the square geometry cases) 
curves in the $(s,t)$-plane,
corresponding to a subdiffusive spreading $\langle s^2\rangle\sim t^b$ 
with an exponent $b \sim 0.5$. 
For the disorder case (with square geometry) the corresponding curves 
of Fig.~\ref{fig4} are also in a qualitative agreement with the reduced 
exponent $b\sim 0.2$ found above by the fit of $\langle r^2\rangle$. 
Concerning the rectangular geometries the curves visible in Fig.~\ref{fig4} 
show saturation also in agreement with Fig.~\ref{fig2} even though 
for the quasiperiodic potential the tails of the distribution (visible 
by light blue zones) still continue 
to increase which is also quite in agreement with the bottom panel 
of Fig.~\ref{fig3}. 

\begin{figure}
\begin{center}
\includegraphics[width=0.48\textwidth]{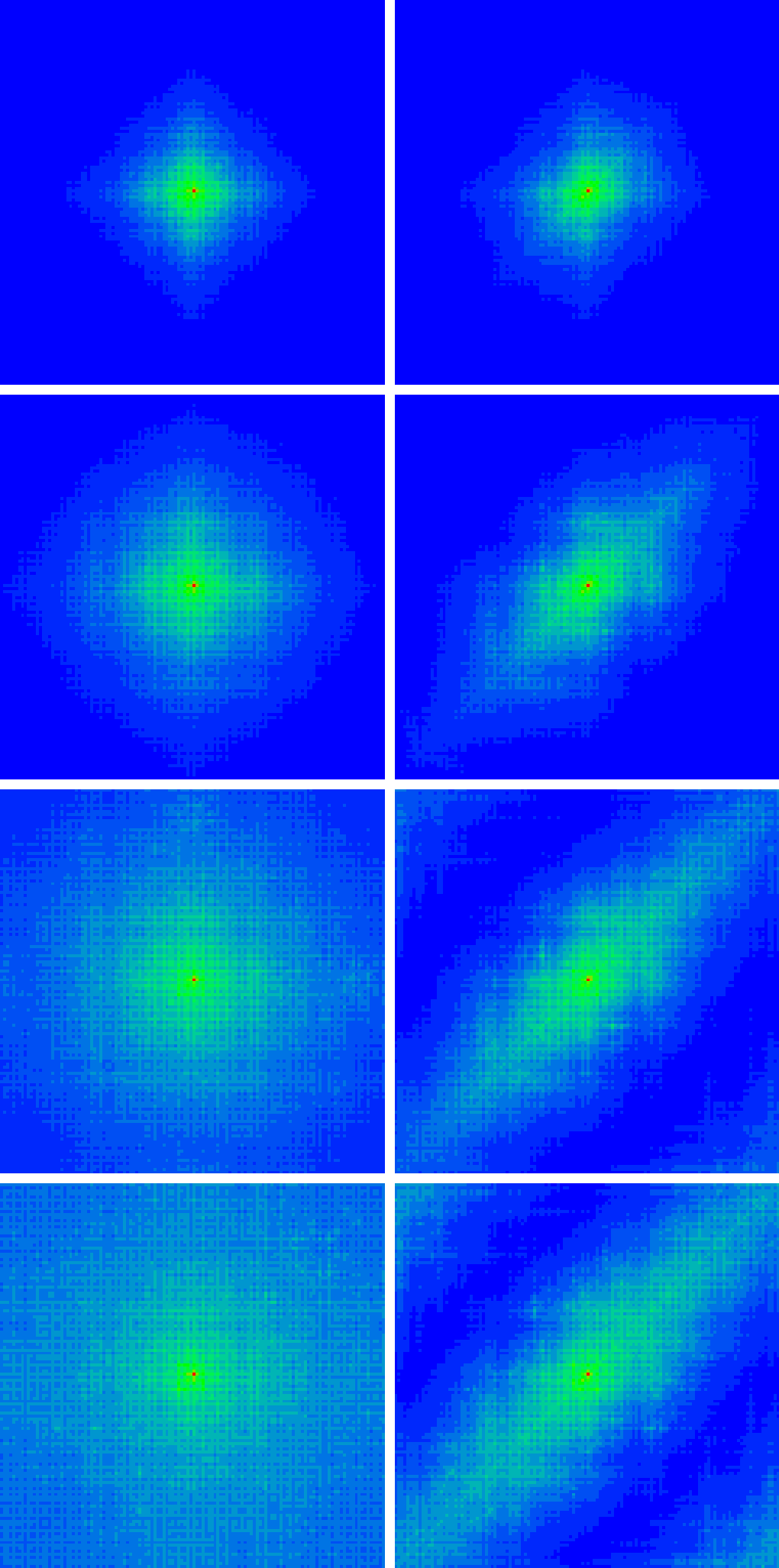}
\caption{Density plot for the densities $\rho_1(x,y)$ (left column) and 
$\rho_{xx}(x_1,x_2)$ (right column) with $x$ (or $x_1$) for the 
horizontal axis and $y$ (or $x_2$) for the vertical axis. All 
panels correspond to $U=3.5$, $U_R=1$ and the square geometry $N_1=N_2=128$ 
with the quasiperiodic potential ($\lambda=2.5$). The different rows 
correspond to the iteration time $t=100$ (first row), $t=1000$ 
(second row), $t=10000$ (third row) and $t=100000$ (fourth row). }
\label{fig5}
\end{center}
\end{figure}

\begin{figure}
\begin{center}
\includegraphics[width=0.48\textwidth]{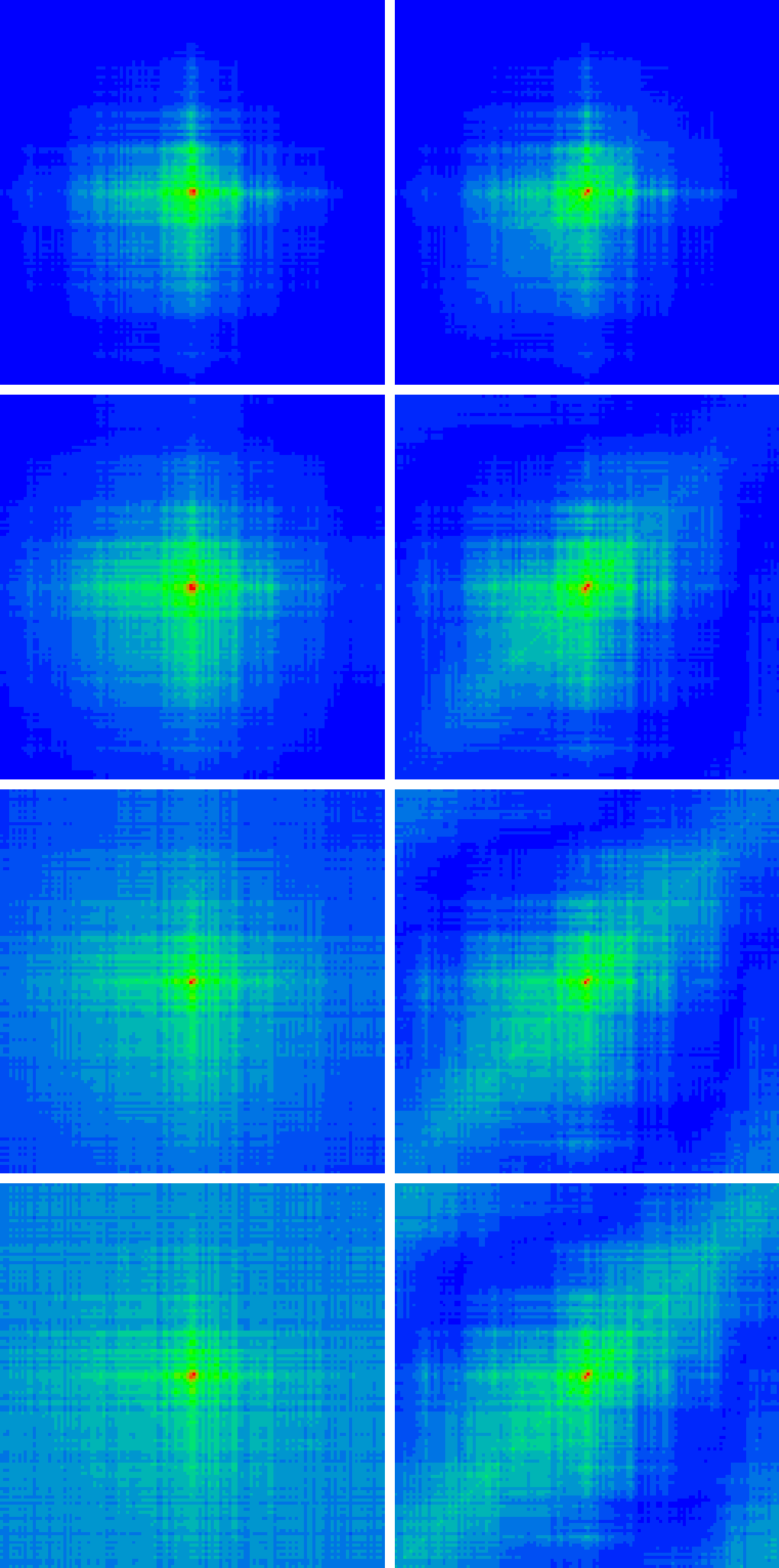}
\caption{The same as Fig.~\ref{fig5} but for the disorder 
potential ($W=5$) and all other parameters identical as in Fig. \ref{fig5}.}
\label{fig6}
\end{center}
\end{figure}

\begin{figure}
\begin{center}
\includegraphics[width=0.48\textwidth]{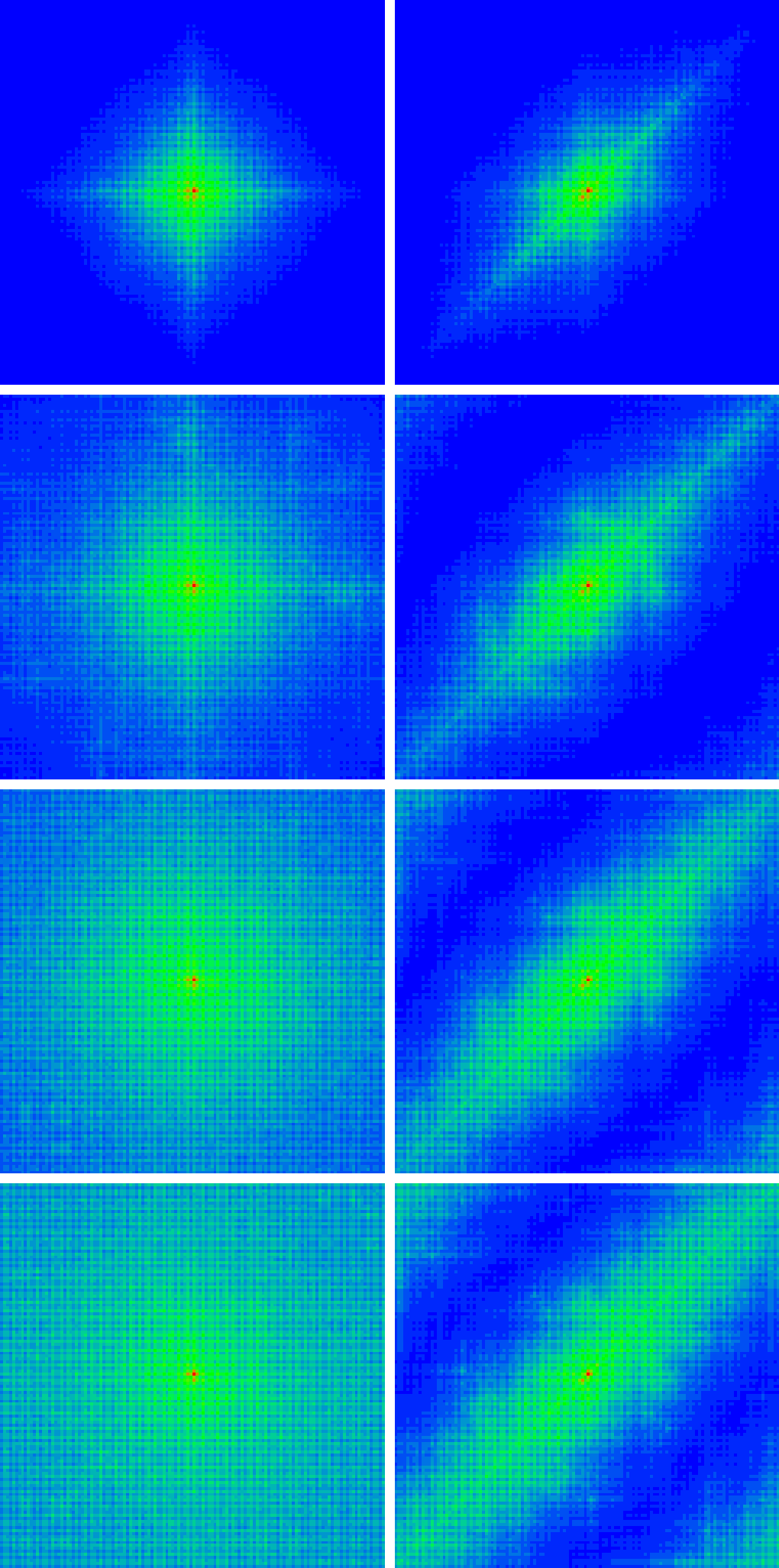}
\caption{The same as Fig.~\ref{fig5} for the quasiperiodic potential 
($\lambda=2.5$) with $U=3.5$, $U_R=2$ and all other parameters 
identical as in Fig.~\ref{fig5}.}
\label{fig7}
\end{center}
\end{figure}

\begin{figure}
\begin{center}
\includegraphics[width=0.48\textwidth]{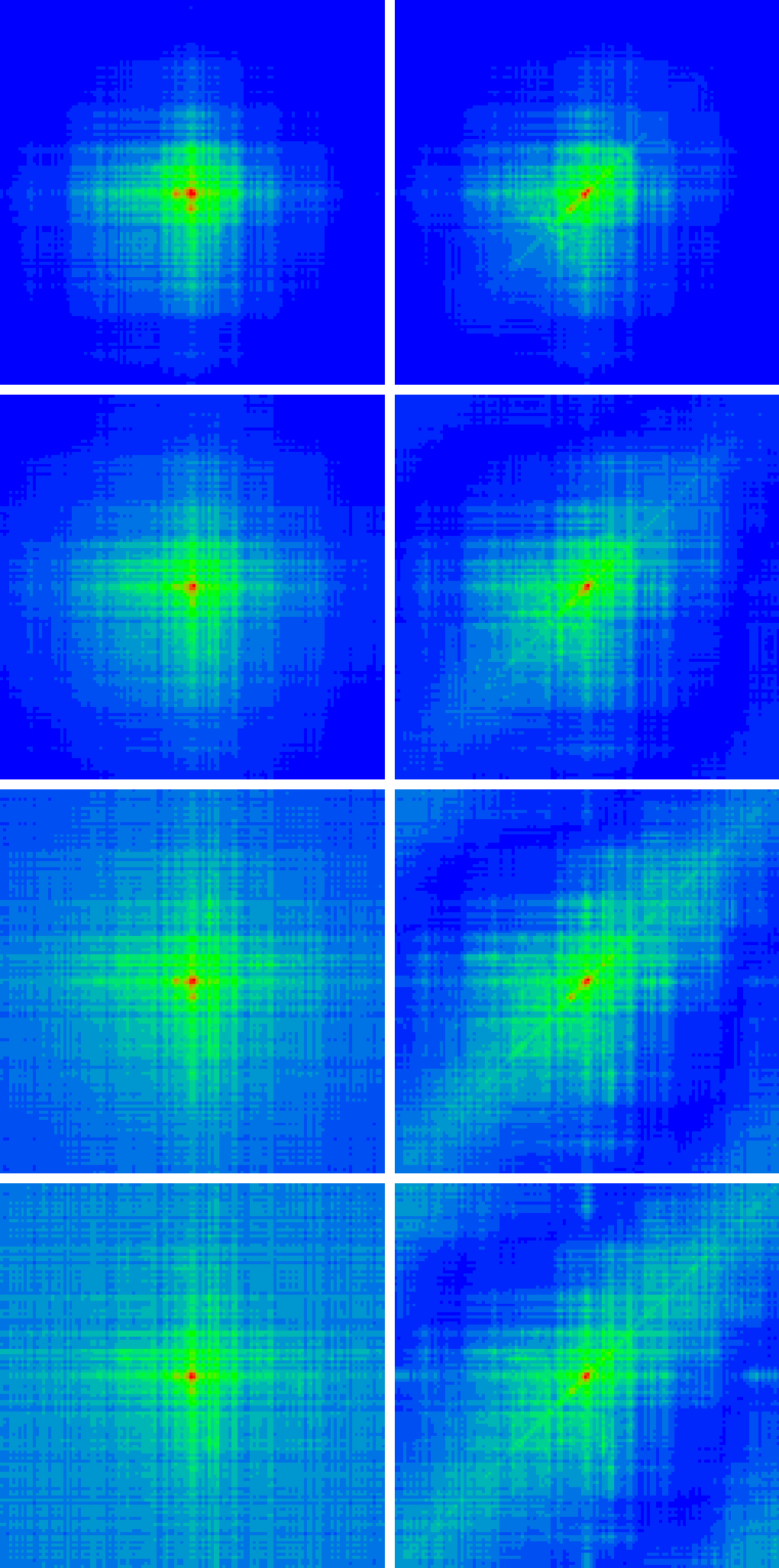}
\caption{The same as Fig.~\ref{fig5} but for the disorder 
potential ($W=5$) with $U=3.5$, $U_R=2$ and all other parameters 
identical as in Fig.~\ref{fig5}.}
\label{fig8}
\end{center}
\end{figure}

The one-particle density $\rho_1(x,y)$ for the square geometry $128\times 128$ 
and $U_R=1$ (or $U_R=2$) is shown at different moments of time
in the left column of Fig.~\ref{fig5} (Fig.~\ref{fig7}) 
for the 2D Harper case and of Fig.~\ref{fig6} (Fig.~\ref{fig8}) 
for the disorder case. The relative distribution of TIP probability in 
the $(x_1,x_2)$-plane, i.~e. the quantity $\rho_{xx}(x_1,x_2)$ defined 
by (\ref{eq_rhoxx}), is shown for the same parameters 
in the right columns of these figures. 

There is a clear spreading of probability in the $(x,y)$-plane 
growing with time.
At largest times $t=10^5$ this spreading starts to saturate due to the 
finite system size and a part of probability returns back due to the 
periodic boundary conditions. This is especially visible in the 
$(x_1,x_2)$-plane with significant contributions in the corners 
$x_1=0, x_2=N_2-1$
and $x_1=N_1-1, x_2=0$ while at shorter times $t \leq 10^4$
the distribution has a well pronounced ``cigar'' shape corresponding
to TIP remaining close to each other. We note that for the Harper case 
the probability distribution inside this cigar
is more homogeneous while for the disorder case there is well visible
cross-structure which we attribute to the fact that we have the same disorder 
structure in $x$ and $y$ directions. In principle, the same is true
for the 2D Harper case but is is possible
that there the localization seems to be better preserved (the cigar is more 
narrow).
Indeed, for the usual 2D uncorrelated disorder the one-particle localization
length at $W=5$ is significantly larger 
as compared to the case of the particular correlated disorder considered here
(see e.g. \cite{lagesbls}). 
In presence of interactions the separability of correlated disorder is broken
that can lead to an additional increase of TIP spearing. Indeed, the width of 
the cigar in the above Figs. is larger for the disorder case. 

The comparison of Figs.~\ref{fig5} and \ref{fig6} also confirms 
the above observation that for $U_R=1$ the quantity $\langle r^2\rangle$ 
is initially (for $t=100$ and $t=1000$) significantly larger for the 
disorder case (Fig.~\ref{fig6}) than for the Harper case (Fig.~\ref{fig5}). 
However, the cross structure visible in Fig.~\ref{fig6} clearly shows that 
this stronger initial delocalization for the disorder case is mostly 
due to stronger individual propagation of one particle in one direction 
and the coherent propagation of TIP sets in at later times while for the 
Harper case the coherent TIP propagation is already important at the 
beginning and dominates the spreading of $\langle r^2\rangle$. We believe 
that the stronger statistical fluctuations of the one-particle 1D 
localization length for the disorder case are partly responsible for 
this observation. We remind that for the Harper 1D model the one-particle 1D 
localization length is really quite constant for all eigenstates while for the 
disorder case there are considerable statistical fluctuations, even for 
one-particle 1D eigenstates of similar energy. 

\begin{figure}
\begin{center}
\includegraphics[width=0.48\textwidth]{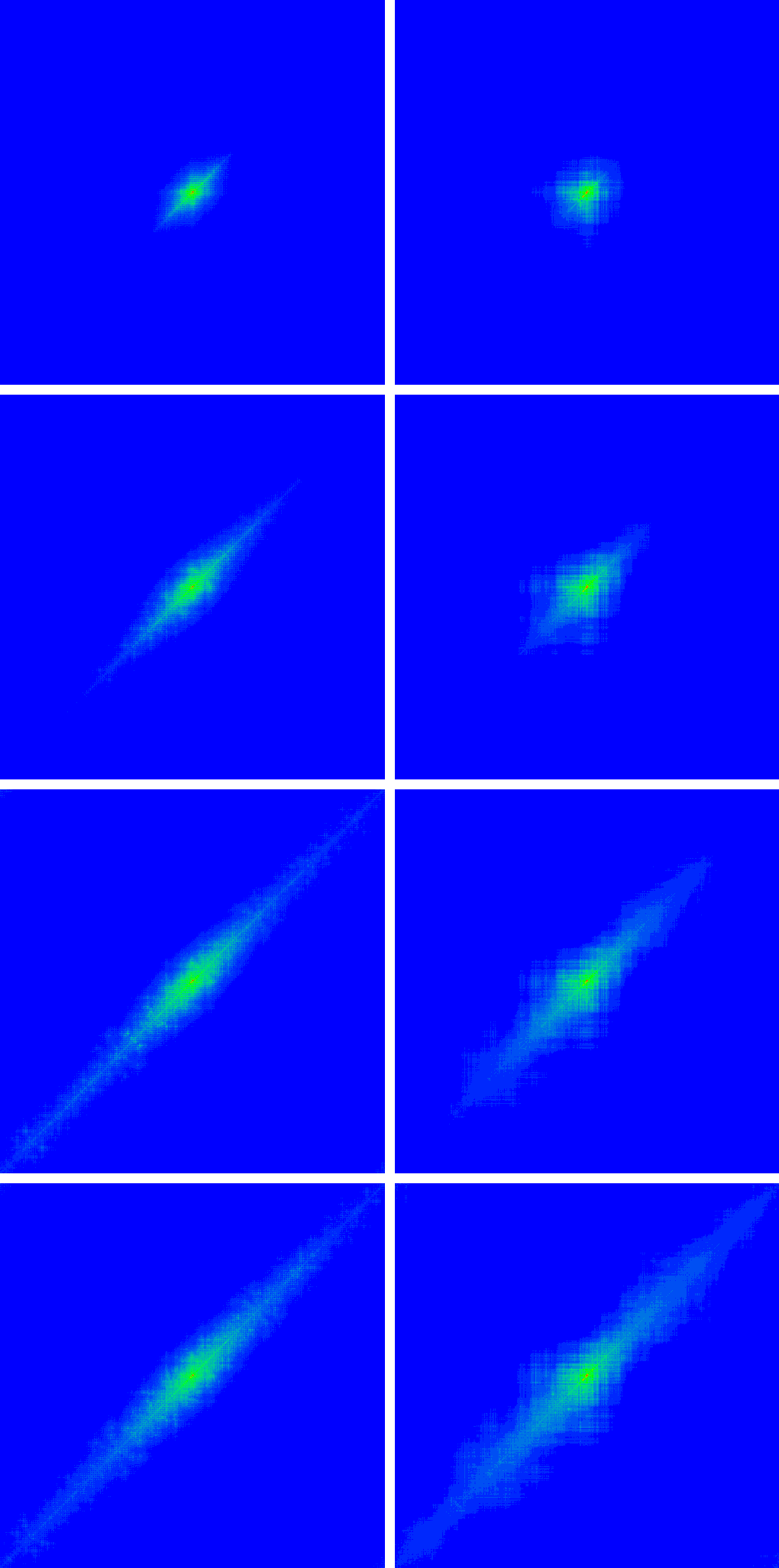}
\caption{Density plot for the density $\rho_{xx}(x_1,x_2)$ with $x_1$ for the 
horizontal axis and $x_2$ for the vertical axis. All 
panels correspond to $U=3.5$, $U_R=2$ and the rectangular geometry 
$N_1=512$, $N_2=8$. The left column corresponds to the quasiperiodic 
potential ($\lambda=2.5$) and the right column to the disorder potential 
($W=5$). The different rows 
correspond to the iteration time $t=100$ (first top row), $t=1000$ 
(second row), $t=10000$ (third row) and $t=100000$ (fourth bottom row). }
\label{fig9}
\end{center}
\end{figure}

The probability distributions for the rectangular geometry are 
shown in Fig.~\ref{fig9}.
In this case the width of the cigar is also smaller in the case of  
the 2D Harper potential as compared to the disorder case. The density at 
$t=10^4$ gives some weak indication on presence of far away probability 
at large $x_1=x_2 \approx N_1$ distances,
which would be expected for ballistic FIKS pairs.
However, the probability there is very small and also
at $t=10^5$ both cases show similar probability profiles
corresponding to localization of the wave packet.

\begin{figure}
\begin{center}
\includegraphics[width=0.48\textwidth]{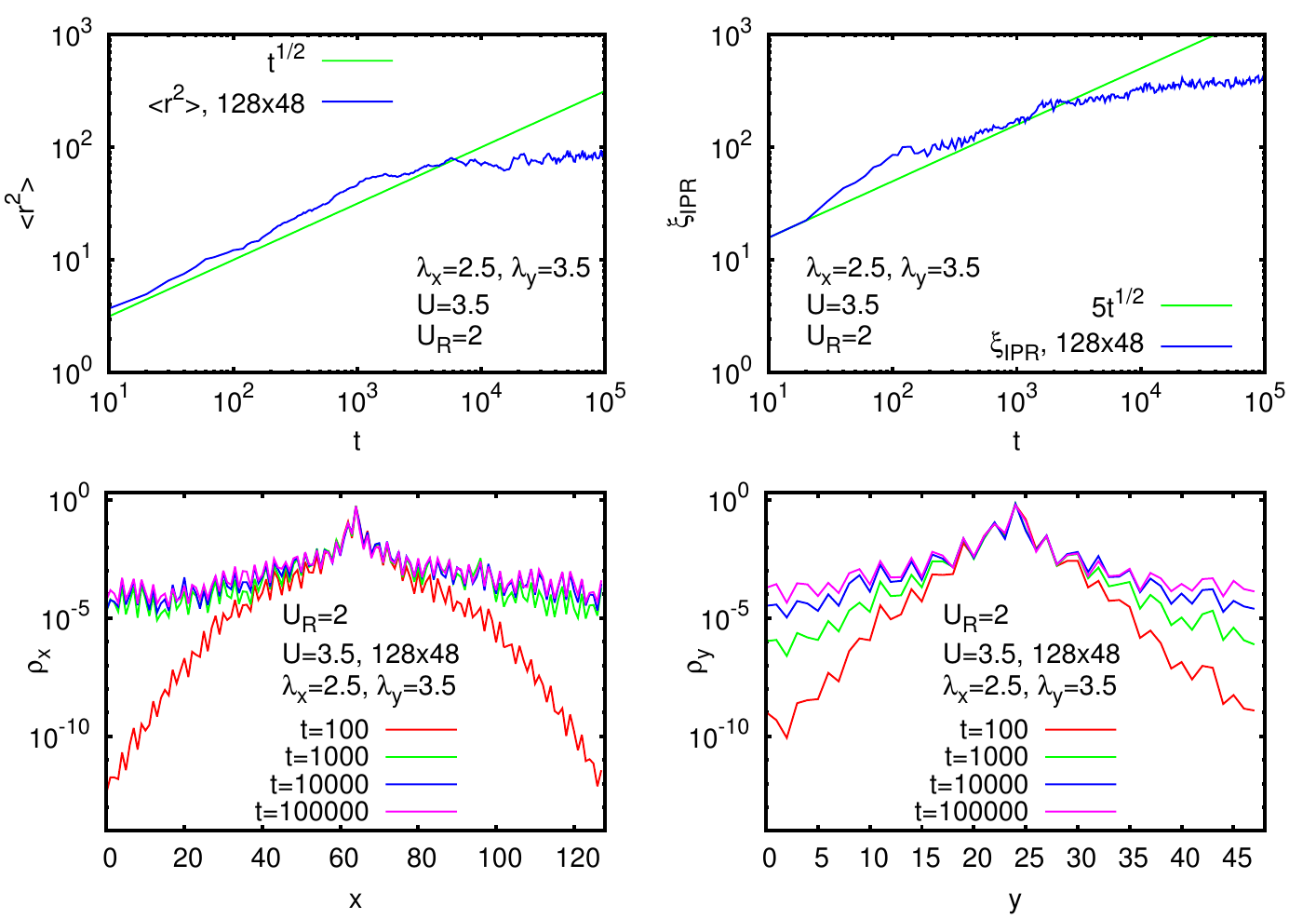}
\caption{Variance $\langle r^2\rangle$ (top left panel) 
and IPR without center $\xi_{\rm IPR}$ (top 
right panel) 
versus iteration time $10\le t\le 10^5$ in a double logarithmic scale 
for the asymmetric case of the quasiperiodic potential 
($\lambda_x=2.5$, $\lambda_y=3.5$) 
with rectangular geometry $N_1=128$, $N_2=48$ and $U=3.5$, $U_R=2$. 
Both quantities are shown by the blue line and the green line shows for 
comparison a power law $\sim t^{1/2}$. 
The bottom left (right) panel shows for the same parameters the 
density $\rho_x(x)$ (or $\rho_y(y)$) versus $x$ (or $y$) 
in a semilogarithmic representation. 
The color labels correspond to different iteration times: 
$t=100$ (red curve), $t=1000$ (green curve), $t=10000$ (blue curve), 
$t=100000$ (pink curve). }
\label{fig10}
\end{center}
\end{figure}

Finally in Fig.~\ref{fig10} we consider an asymmetric case 
of the 2D Harper model with $\lambda_x=2.5$, $\lambda_y=3.5$, 
$N_1=128$ and $N_2=48$. 
Here we have a significantly stronger localization of non-interacting 
particles in the $y$-direction with $\ell_y=1/\log(\lambda_y/2)\approx 1.79$. 
Thus we could expect appearance of 1D ballistic 
FIKS pairs in such a case. However, this scenario is not confirmed 
by the data which still give a subdiffusive spreading
with the fit exponents $b_1=0.563 \pm 0.004$ and $b_2=0.431 \pm 0.016$
for the time range $10 \leq t \leq 1000$ and the power law fits 
$\langle r^2\rangle \propto t^{b_1}$ and 
$\xi_{\rm IPR} \propto t^{b_2}$. The probability distribution
in $x$ becomes rather broad at large times $t=10^5$
and it is possible that even larger system sizes are required to firmly 
state if this subdiffusion continues on longer times. Furthermore 
the density $\rho_y(y)$ does not show a strong localization in the 
$y$-direction in presence of interaction, despite the very small value 
of $\ell_y$, and there are quite large 
tails of $\rho_y(y)$ for $y$ being close to the transversal boundaries. 
Therefore the scenario of an effective $1D$-situation 
in $x$ due to strong $y$-localization does not really happen thus explaining 
that we have no visible indications for FIKS pairs 
in such an asymmetric situation. 

\section{Discussion}
\label{sec4}

We presented here the study of interaction effects in the 2D Harper model
where the two-dimensional quasiperiodic potential is given as 
the sum of two one-dimensional quasiperiodic potentials for the $x$ and 
the $y$ direction. Our results show 
that in this system the interactions induce a subdiffusive
spreading over the whole lattice with the spreading exponent
being approximately $b \approx 0.5$ for the second moment and IPR.
Such a delocalization takes place in the regime when all one-particle 
eigenstates are exponentially localized. 
In this 2D TIP Harper model we do not find signs of ballistic FIKS pairs, 
which are well visible for the 1D TIP Harper case \cite{flach,fiks1d}.

It is possible that the physical reason of absence of FIKS pairs in 2D Harper model
is related to the fact that for TIP in 2D we have a much more dense
spectrum of non-interacting eigenstates [see e.g. Eg.(29) in \cite{fiks1d}
where the indexes $m_1, m_2$ of non-interacting eigenstates of two particles 
now become vectors in 2D]. Due to this there are practically no well separated energy bands
typical for the one-particle 1D Harper model and thus
there is little chance to have an effective Aubry-Andr\'e Hamiltonian
with $\lambda_{\rm eff}$ and the interaction induced hopping matrix 
elements $t_{\rm eff}$
generating a metallic phase with $\lambda_{\rm eff} < 2 t_{\rm eff}$. 
Of course, there is still a possibility
that we missed some FIKS cases at specific $U$ values but
for all studied cases of TIP in the 2D Harper model we find a
subdiffusive spreading being qualitatively different from 
the FIKS effect in the 1D Harper case.
For a rectangular geometry 
with a narrow size band in one direction 
we even obtain a localization of TIP spreading.

When the quasi-periodic potential is replaced by a disorder potential 
of the particular form (\ref{eq_pot_form}) we also find a subdiffusive 
spreading but with 
a smaller exponent $b \approx 0.25$ (on available time range and system size).
In principle, for TIP in the 2D disorder potential we
expect to have localized states for short range interactions 
\cite{imry,dlsmoriond,dlscoulomb}. However, here we consider 
a particular correlated disorder (with a potential being a sum of two 
one-dimensional potentials in $x$ and $y$) 
and in such a case the one-particle localization length
at $W=5$ ($\ell_1 \approx \xi \approx 5$)
is significantly smaller than for the usual 2D disorder potential 
(see e.g. \cite{lagesbls} with $\xi \approx 150$). We think that in 
presence of interactions and sufficient iteration times 
such correlations of disorder are suppressed
and we have a situation similar
to the TIP case of the usual 2D Anderson model  
where at $W=5$ the one-particle localization length $\ell_1$ is rather large 
and thus the TIP localization length $\ell_2$, 
expected to be an exponent of $\ell_1$ \cite{imry,dlsmoriond},
is also very large ($\ln \ell_2 \sim l_1$) 
and is not reachable at time scales and system sizes used 
in our studies. In any case the smaller value of $b \approx 0.25$ for 
the disorder case, compared to the 2D Harper case with $b \approx 0.5$,
indicates that some residual effects of FIKS pairs give a stronger
delocalization of TIP for the 2D Harper model.

It is interesting to note that a somewhat similar sub\-diffusive spreading
appears in the 2D Anderson model with a mean field type
nonlinearity (see e.g. \cite{garcia}). However, there the value
of the spreading exponent $b \approx 0.25$ is smaller
(the value $b \approx 0.5$  found here is more similar to the
1D Anderson model with nonlinearity studied in \cite{danse,flachdanse}). 
However, the physical origin of a certain similarity
of these nonlinear mean-field models with the TIP case studied here 
remains unclear since here we have a linear Schr\"odinger equation
while the models of \cite{danse,garcia,flachdanse} are described by
classical nonlinear equations (second quantization is absent).

We think that the 2D TIP Harper model provides us new 
interesting results with subdiffusive spreading 
induced by interactions.
This model rises new challenges for advanced mathematical 
methods developed for quasiperiodic Schr\"odinger operators
\cite{lana2,lana3}. It is also accessible to 
experimental investigations with ultracold atoms
in 2D quasiperiodic optical lattices which 
can be now built experimentally \cite{bloch2d}.
Thus we hope that the TIP problem in 1D and 2D Harper models will
attract further detailed theoretical and experimental investigations.

This work was granted access to the HPC resources of 
CALMIP (Toulouse) under the allocation 2015-P0110. 

%\vskip -1.5cm
%%% One dummy reference 


\begin{thebibliography}{99}
\bibitem{U_boundary} In view of the periodic boundary conditions 
the condition $|x_1-x_2|<U_R$ (or $|y_1-y_2|<U_R$) 
is understood to be true also for the case 
$N_1-|x_1-x_2|<U_R$ ($N_2-|y_1-y_2|<U_R$), 
i.~e. if $x_1$ ($y_1$) is close to one boundary and $x_2$ ($y_2$) 
to the other boundary.
\end{thebibliography}
\end{document}